\documentclass{jpp}

\usepackage{bm}
\usepackage{graphicx}
\usepackage{natbib}
\usepackage{soul}

\ifCUPmtlplainloaded \else
  \checkfont{eurm10}
  \iffontfound
    \IfFileExists{upmath.sty}
      {\typeout{^^JFound AMS Euler Roman fonts on the system,
                   using the 'upmath' package.^^J}%
       \usepackage{upmath}}
      {\typeout{^^JFound AMS Euler Roman fonts on the system, but you
                   dont seem to have the}%
       \typeout{'upmath' package installed. JPP.cls can take advantage
                 of these fonts, if you use 'upmath' package.^^J}%
      }
  \else
  \fi
\fi

\ifCUPmtlplainloaded \else
  \checkfont{msam10}
  \iffontfound
    \IfFileExists{amssymb.sty}
      {\typeout{^^JFound AMS Symbol fonts on the system, using the
                'amssymb' package.^^J}%
       \usepackage{amssymb}%
         \let\leq=\leqslant
         
      }{}
  \fi
\fi

\ifCUPmtlplainloaded \else
  \IfFileExists{amsbsy.sty}
    {\typeout{^^JFound the 'amsbsy' package on the system, using it.^^J}%
     \usepackage{amsbsy}}
    {}
\fi

\newsavebox{\astrutbox}
\sbox{\astrutbox}{\rule[-5pt]{0pt}{20pt}}

\usepackage{color}

\title[Flow dynamics and magnetic induction in VKP]{Flow dynamics and magnetic induction in the von-K\'arm\'an plasma experiment}

\author[Plihon, Bousselin, Palermo {\it et al.}]%
{
N. PLIHON$^1$%
  \thanks{Email address for correspondence: nicolas.plihon@ens-lyon.fr},\ns
G. BOUSSELIN$^1$,
\ns
F. PALERMO$^{1,2}$,
\ns 
J. MORALES$^2$,
\ns
W.J.T. BOS$^2$,
\ns 
F. GODEFERD$^2$,
\ns 
M. BOURGOIN$^1$,
\ns
J.-F. PINTON$^1$,
\ns
M. MOULIN$^1$
\and 
\ns
A. AANESLAND$^3$}
\affiliation{$^1$Laboratoire de Physique, \'Ecole Normale Sup\'erieure de Lyon, CNRS \& Universit\'e de Lyon, 46 all\'ee dÕItalie, 69364 Lyon Cedex 07, France\\[\affilskip]
$^2$LMFA-CNRS, Universit\'e de Lyon, \'Ecole Centrale de Lyon, 69134 Ecully, France\\[\affilskip]
$^3$Laboratoire de Physique des Plasmas (CNRS, \'Ecole Polytechnique, Sorbonne Universit\'es, UPMC Univ Paris 06, Univ Paris-Sud),\' Ecole Polytechnique, 91128 Palaiseau, France}

\pubyear{}
\volume{}
\pagerange{}
\date{?; revised ?; accepted ?. - To be entered by editorial office}
\begin{document}

\maketitle

\begin{abstract}
The von-K\'arm\'an plasma experiment is a novel versatile experimental device designed to explore the dynamics of basic magnetic induction processes and the dynamics of flows driven in weakly magnetized plasmas. A high-density plasma column ($10^{16}$--$10^{19}$  particles.m$^{-3}$) is created by two radio-frequency plasma sources located at each end of a 1 m long linear device. Flows are driven through $\bm J \times \bm B$ azimuthal torques created from independently controlled emissive cathodes. The device has been designed such that magnetic induction processes and turbulent plasma dynamics can be studied from a variety of time-averaged axisymmetric flows in a cylinder.  MHD simulations implementing volume-penalization support the experimental development to design the most efficient flow-driving schemes and understand the flow dynamics. Preliminary experimental results show that a rotating motion of up to nearly 1 km/s is controlled by the $\bm J \times \bm B$ azimuthal torque.
\end{abstract}


\section{Introduction}

The coupling between velocity and magnetic fields in electrically conducting fluids is ubiquitous in nature and very important in key applications. Magnetic fields of astrophysical bodies originate from the dynamo instability, in a process where kinetic energy is converted into magnetic energy~\citep{Moffatt}. In the immediate vicinity of the Earth, the interaction of the energetic particles of the solar wind~\citep{Goldstein} with the Earth's magnetosphere leads to spectacular polar aurorae.  In most of the situations, fluctuations of the physical parameters play a crucial role in the dynamics of the systems and make them unpredictable (as for instance space weather or magnetic storms forecasts) or difficult to control (e.g. turbulence in thermonuclear fusion plasmas).

Within the magnetohydrodynamic (MHD) frame of description, the joint $(\bm{u},\bm{B})$ dynamics is described by the Navier-Stokes equation for the velocity field $\bm{u}$  and the induction equation for the magnetic field $\bm{B}$:

\begin{eqnarray}
 \rho\left(\frac{\partial \bm u}{\partial t}+\bm u\cdot \nabla \bm u\right)=-\nabla p +\eta_u \Delta \bm u+\bm J\times\bm B,\label{eq:u}\\
 \frac{\partial \bm B}{\partial t}=\nabla \times (\bm u \times \bm B)+\eta_B \Delta \bm B\label{eq:B}
\end{eqnarray} 

\noindent where $\eta_B$ and $\eta_u$ are respectively the magnetic and dynamic viscosities, $\rho$ is the fluid density, $p$ is the fluid pressure, $\bm{J}$ is the current density.
From these equations, two independent dimensionless numbers control the dynamics of the system:
 \begin{itemize}
\item The magnetic Reynolds number Rm, compares the amplitudes of the
induction term to the dissipation term and describes the importance of non-linear effects in the induction equation. Rm can be defined as a function of a characteristic velocity $U$ and a characteristic length scale $L$ of the flow as ${\rm Rm} =UL/\eta_B$. One should note that there are usually several possible choices for the characteristic length scale and velocity, leading to different scaling of this control parameter. One should also note that the magnetic viscosity $\eta_B$ depends on the electrical conductivity $\sigma$ of the conducting medium and on the magnetic permeability $\mu_0$ as $\eta_B = (\mu_0 \sigma)^{-1}$, and thus ${\rm Rm} =\mu_0\sigma UL$. 
\item The magnetic Prandtl number Pm, defined as the ratio of kinematic viscosity over the magnetic viscosity -- as $ {\rm Pm} = \displaystyle\frac{\eta_u}{\rho\eta_B} $--, compares the diffusive time or length scales of the velocity and magnetic fields and depends on the physical properties of the fluid. 
\end{itemize}
A third relevant dimensionless number is the kinetic Reynolds number Re, which is defined as the ratio of inertia over viscous effects in the fluid, and is simply given by the ratio Rm/Pm. Re characterizes the flow regime, possibly turbulent. The intensity of the flow drive is characterized by the ratio of the strength of the Lorentz force over that of inertia or viscous dissipation, depending on the flow regime. The complexity of the joint dynamics of velocity and magnetic fields arises from non-linear effects in the induction and/or Navier-Stokes equations, at large values of the kinetic Reynolds number Re and/or magnetic Reynolds number Rm. Investigations of the dynamics of the velocity field with fixed magnetic field or of the dynamics of the magnetic field in a prescribed velocity field have been extensive. While it is well known that keeping the fields separate is misleading, these studies have been instrumental for studying the elementary processes of the full non-linear problem. 
Liquid metal flows have been widely used for such studies, especially flows created by  counter-rotating disks or impellers, and named von-K\'arm\'an (VK) swirling flows after~\cite{Zandbergen}. VK flows have attracted much attention over the last two decades in the fluid mechanics / turbulence communities due to their unique ability to generate a high level of turbulence in confined geometry~\citep[see for instance][]{DouadyPRL91}. 
A detailed investigation of the transition to turbulence in VK water flows may be found in~\cite{RaveletJFM2008}. In fully developed turbulent regimes, VK flows were shown to sustain large-scale instabilities~\citep{delaTorrePRL2007} and to display multistability~\citep{RaveletPRL2004}. A sketch of a time-averaged VK flow created in a cylinder by the counter-rotation of two impellers fitted with blades is given in figure~\ref{figure:VKschematic} (upper left panel). It consists of two toroidal cells rotating in
opposite directions and two poloidal cells due to the centrifugal forces in the vicinity of the impellers, as sketched in  (in a cylinder the axis of symmetry being in the axial direction, the toroidal direction is the azimuthal --- or orthoradial --- direction, while the poloidal plane is the $(r,z)$ plane). This flow is a cylindrical equivalent of the $s_2t_2$ flow introduced by~\cite{DudleyJames} ($s$ denoting the poloidal cells and $t$ the toroidal cells) and possesses both large scale kinetic shear and helicity which makes it attractive for turbulent MHD studies.  
These studies may be divided in two categories:
\begin{itemize}
\item Magnetic induction studies. In this context, work has aimed at uncovering efficient induction processes that could co-operate towards dynamo generation. In these studies an external field $B^A$ is applied, and one analyses magnetic response, i.e. the induced field $B^I$, for values of Rm typically below 10 (the stable fixed point of the induction equation being then $B=0$ in the absence of an externally applied field). Details of the studies carried out in VK flows may be found in~\cite{VerhilleSSR2010}, with emphasis on the effect of differential rotation (the $\omega$-effect), of kinetic helicity (the Parker-effect) or on magnetic expulsion from vortex motions.
\item Dynamo studies. The dynamo instability is observed when induced currents from flow motion overcome resistive Joule dissipation, requiring the magnetic Reynolds number to exceed a critical value ${\rm Rm}_c$ (typically of order 100 in VK flows). This is however not a sufficient condition: the details of the flow have to promote an efficient feedback of magnetic induction mechanisms to lead to magnetic field amplification and growth of magnetic energy. Saturation of the magneticÊ energy is reached because of flowÊ modificationsÊ generated byÊ the Lorentz force. Several experimental attempts have been carried using VK flows over the last decade~\citep{Peffley,Spence,P5}. Dynamo action has been reached in the VKS experiment~\citep{P5} when the flow is driven by impellers made from high permeability materials, via a supercritical bifurcation with a critical value of the control parameter ${\rm Rm}_c\sim 44$. Statistically stationary dynamos have been observed, as well as a variety of dynamical regimes (reversals, bursts or oscillations). Recently, a semi-synthetic dynamo was also developed at lower values of Rm, which, for instance, allows to reach strongly saturated regimes~\citep{Bourgoin2006,Miralles2014}.
\end{itemize}
It should be emphasized that magnetic induction and dynamo studies were also obtained with flows driven by differentially rotating impellers. In these regimes, the number of toroidal and poloidal cells can be varied, as well as their relative size. This has strong implications on the magnetic response of the flow~\citep{VerhilleSSR2010}. Figure~\ref{figure:VKschematic} sketches one of these asymmetric situations, where only one impeller rotates, leading to a time-averaged $s_1t_1$ flow. These studies have been carried out using liquid metal as a conducting medium. In liquid metals Pm is of the order of $10^{-6}$, and non-linear effects in the induction equation (which require high values of Rm) are only obtained for very large values of the kinetic Reynolds number Re, i.e. for highly turbulent flows. This leads to two main limitations. The first one concerns the maximum achievable Rm, which scales with mechanical power $P$ as $P^{1/3}$. Typical Rm values around 100 require 100's of kW of mechanical power when using liquid sodium (Rm values around 1000 would require 100's of MW !). The second limitation deals with the high level of fluctuations, which may inhibit induction processes (and increase the threshold for onset of dynamo action)~\citep{Rahbarnia}. Using fluids with variable Pm would extend the range of operational parameters. By varying the value of Pm, one tunes the non-linear coupling between the ($\bm u, \bm B$) fields.  While in liquid metals, non-linear effects in the induction equation are only obtained in highly turbulent flows, the situation might be more versatile in plasmas, where the value of Pm depends on transport coefficients -- electrical conductivity and kinematic viscosity -- ~\citep{Braginskii1965}, depending on the plasma parameters through the collision rates between particles. Pm values in the range $10^{-7}$ to 10 are accessible to state-of-the-art experimental plasma devices. The group of Pr Forest at UW Madison first proposed to use plasmas in a large-scale experiment to study dynamo action at large Pm value~\citep{Spence2009}. We present in this article an ongoing experimental project aiming at developing a von-K\'arm\'an type flow in a weakly magnetized plasma for investigations of basic magnetic induction processes in a plasma, with variable Pm values, at Rm values of order 10. Flows will be forced in a linear device by a $\bm J \times \bm B$ torque using emissive cathodes. This project is complementary to the ongoing studies at UW Madison on stirring large-scale unmagnetized plasmas in the Plasma Couette Experiment~\citep{Collins2012} and in the Madison Plasma Dynamo Experiment~\citep{Cooper2014} where a $\bm J \times \bm B$ torque with a multicusp confining magnetic field is used.
 The scientific aim of our project is three-fold:
\begin{itemize}
\item To extend the studies on turbulent MHD processes and dynamo instability from very low Pm value ($10^{-6}$) to larger values (up to 10) using plasma as a conducting fluid. Deviations from the MHD frame (two fluid effects, Hall term) will be detailed.  The wide range of accessible Pm values gives a unique opportunity to bridge regimes obtained experimentally in liquid metals to regimes currently reached in numerical simulations~\citep{Ponty}.
\item To investigate the dynamics of plasma flows and plasma parameters fluctuations in the presence of large-scale driven flows (rotation, axially sheared flows)
\item The above points require the development of ad-hoc plasma acceleration schemes and plasma flow control. A detailed investigation of transport processes in weakly magnetized, partially ionized plasmas and the influence of Pm will also be carried out.
\end{itemize}

Creating plasma flows in magnetized plasmas has been achieved in numerous linear and toroidal magnetized experiments~\citep{Schaffner,Zhou,Klinger,Annaratone,Teodorescu,Wallace,Fredriksen}. In most of these investigations, rotation of the plasma column is achieved by biasing cold limiters, rings or portions of the vacuum vessel. In these studies, the influence of radial plasma flow shear on interchange instability~\citep{Teodorescu}, drift wave instabilities~\citep{Klinger}, turbulent transport~\citep{Schaffner,Zhou,Wallace}, plasma confinement~\citep{Fredriksen} or radial transport~\citep{Annaratone,Oldenburger}  has been detailed. Our flow driving scheme will allow to drive plasma rotation in both directions and to extend some of the previous studies in cases with an axial shear of the rotation of the plasma column. 
The project has recently started and is in its initial stage; this article is thus rather written  in the spirit of a proof of concept paper than as a review paper. The experimental device is described in detail in section~\ref{sec:2}, together with the expected capabilities. In this section, both the current stage of the project and the expected final state are discussed. The addressed scientific questions are then detailed in section~\ref{sec:3}. Preliminary results are discussed in section~\ref{sec:4}. 


\section{Machine description and capabilities}\label{sec:2}

\subsection{The von-K\'arm\'an plasma experiment}

\begin{figure}
\includegraphics[width = \columnwidth]{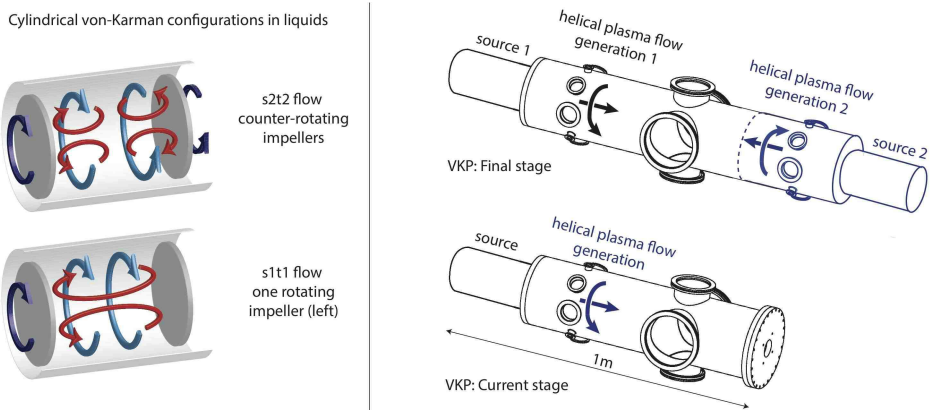}
\caption{[color online](Left) Schematic of VK flows in conventional with (top) counter-rotating impellers and (bottom) only one rotating impeller (left one). Toroidal flows are sketched with blue arrows, poloidal motions with red arrows. (Right, top) Final stage of the VKP project, showing the two counter rotating plasma cells and (bottom) current stage device.}
\label{figure:VKschematic}
\end{figure}

\begin{figure}
\includegraphics[width = \columnwidth]{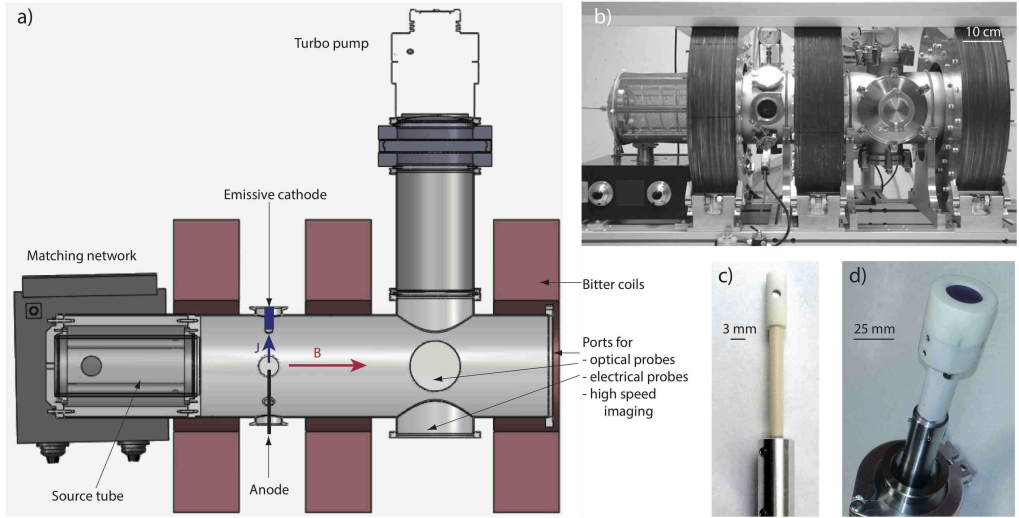}
\caption{(a) Schematic of the current setup, showing the driving electrode configuration close to the radio-frequency source. (b) Photograph of the experiment (c) two-face Tantalum Mach probe and (d) LaB$_6$ emissive cathode.}
\label{figure:Expschematic}
\end{figure}

The final projected experimental setup is sketched in figure~\ref{figure:VKschematic} (upper right panel) and consists of a cylindrical vacuum vessel with plasma sources at each end  of the cylinder. Independent control of plasma parameters and flow drive is achieved from two independent acceleration regions located close to the plasma sources. This scheme will allow to control independently azimuthal plasma rotation and axial flow, leading to axial shear of the azimuthal velocity as well as axial compression.
The current device consists of a single cell, i.e. one plasma source and one acceleration region, as shown on the lower right panel of figure~\ref{figure:VKschematic}, together with a detailed design and photograph of the experiment in figure~\ref{figure:Expschematic}.

The plasma source consists of a cylindrical pyrex tube of internal diameter 110 mm around which a 3 turns radio-frequency coil is fed by a 13.56 MHz radio-frequency power generator (up to 2 kW) through a close-coupled L-type matching network. The actual design promotes an inductive coupling to ease the development stage of flow drive (as described below). In the near future, we plan to develop an optimised helicon antenna will be developed in order to increase the ionisation fraction up to nearly 100\%~\citep{Boswell}. A turbo-molecular pump routinely maintains a
base pressure of $10^{-5}$~Pa, while plasma is created using noble gases (Ar, He, Kr or Ne  allowing to modify the ion mass) in the range $10^{-2}$--30~Pa. The vessel has a diameter of 200~mm and a length of 700~mm. Bitter coils provide an axial magnetic field up to 0.2~T. Typical plasma parameters of the current stage device and the final stage projected setup are presented in table~\ref{tab:plasmaparam}.
The radio-frequency source, together with the weak confining axial magnetic field, creates a plasma column that extends along the axial direction. A typical radial profile of the plasma density and electron temperature is given in figure~\ref{Fig:expresults}(a), which shows a 7 cm diameter plasma column, with a clear density and electron temperature gradient at the edge of the plasma column.

The flow driving region is adjacent to the source region. The current scheme, under optimisation, is based on large scale $\bm J \times \bm B$ azimuthal torque from a radial current and an axial magnetic field. The radial current is driven and controlled by an emissive cathode inserted in one of the radial ports in the acceleration region, negatively biased relative to an anode. Our initial cathode design was based on 0.25 mm diameter Thoriated Tungsten filaments Joule-heated up to above 2700 K (as measured from a Raytech Marathon pyrometer). Preliminary results obtained with this configuration are presented in the next section. This design suffers from two main limitations: a moderate emitted current (up to a few Amperes) and tungsten sputtering, spoiling electrostatic probes. A new design has been developed with a 25 mm diameter emissive LaB$_6$  disk (see figure~\ref{figure:Expschematic}) that should increase the emitted current and lower metallic deposition. This flow drive will control the azimuthal flows of  the plasma column, independently of the plasma creation, controlled by the radio-frequency source. In the near future, the development of a ring emissive cathode --- in order to recover the azimuthal symmetry --- or flow drive from a multicusp magnetic confinement as demonstrated in~\cite{Katz} may also be envisioned.
The current device, consisting of one-cell of the planned von-K\'arm\'an flow  allows to drive a $s_1t_1$ type flow. The final device will drive $s_2t_2$ type flows, with a strong axial shear of the azimuthal velocity; and depending on the relative value of the azimuthal rotation, the transition between the two types of flows will be studied. Driving $s_2t_2$ flows with a homogenous external axial magnetic field will require current flowing radially outwards at one end and flowing radially inwards at the other end; this will be achieved from emissive cathodes located on opposite azimuthal angles. An alternative scheme will promote a cusp field in the central region with radial emitted currents flowing in the same direction (either outwardly or inwardly) at each end of the cylinder. While the first scheme could be obtained with only one plasma source, the second one requires two independent sources. Note that in this configuration, axial flows could be driven from the magnetic expansion of the plasma~\citep{Charles,Plihon}. Alternative schemes such as single grid acceleration of the plasma~\citep{Dudin} may also be implemented to independently control the ratio of the poloidal ($s$) and toroidal ($t$) flows.
Finally, magnetic coils creating a small amplitude transverse magnetic field  will also be installed for magnetic induction studies.

\subsection{Plasma parameters and capabilities}
Table~\ref{tab:plasmaparam} gives the typical plasma parameters obtained in the current device and those expected in the final device. 
In this table, the flow velocity is specified as the azimuthal velocity driven from the $\mathbf{J}\times\mathbf{B}$ torque created by current flowing from the emissive cathode to the positively biased anode. For the range of plasma parameters given in table~\ref{tab:plasmaparam} and for neutral pressure in the range 0.01 to 1 Pa, the electrons are weakly magnetized while the ions are not. Ion collision frequencies being larger or of the order of the  ion gyrofrequency, the collisions prevent the gyromotion, and the ratio $\omega_{ci}/\nu$ is of order one ($\omega_{ci}$ being the ion gyrofrequency and $\nu$ the ion collision frequency). Thus the Pedersen and Hall conductivities are non vanishing --- and typically of the order of the parallel conductivity. The plasma collisionality thus allows for an efficient flow drive from the azimuthal $\bm J \times \bm B$ torque. Flow drive in other experiments with similar plasma parameters also rely on  finite perpendicular plasma conductivity~\citep{Collins2012,Cooper2014,Zhou,Klinger,Annaratone}.

One of the assets of the proposed setup lies in the possibility to reverse the rotation direction while keeping the axial magnetic field direction constant by permutation of the emissive cathode and  anode locations (the configuration cathode at the edge / anode in the center provides an outward radial current while the opposite configuration drives an inward radial current). This gives an extra free parameter as compared to schemes based on biasing cold cathodes, limiters or portions of the vacuum vessel. Preliminary results, shown below, have demonstrated plasma flows around 1 km.s$^{-1}$, and optimized configurations are expected to reach plasma flows in the range 10 km.s$^{-1}$ in Argon and 20 km.s$^{-1}$ in Helium. The intensity of the flow is controlled by the amplitude of the emitted current from the cathode independently of the plasma density, which is controlled by the radio-frequency power coupled to the plasma.

The scientific goals of the project have been detailed above and focus on the influence of the magnetic Reynolds number Rm and the magnetic Prandtl number Pm on magnetic induction processes and plasma flow characteristics. In weakly magnetised, partially ionised plasmas the dynamics of the particles is anisotropic and transport coefficients are described by tensors. This is indeed the case for the plasma conductivity, the plasma viscosity as well as for the neutral drag. A precise experimental determination of the various terms of these tensors is indeed one of the questions that will be addressed, and there are no simple expressions for Rm and Pm. However, for the sake of simplicity, we will restrict ourselves here to the case of parallel coefficients, along the axis, (or equivalently of an unmagnetized plasma) to express scalings and typical values of the dimensionless parameters expected in our setup. According to~\cite{Braginskii1965}, the magnetic Reynolds and Prandtl numbers in a plasma with singly charged ions read

\begin{eqnarray}
{\rm Rm} = 1.6\ T_{e,\rm eV}^{3/2}U_{\rm km/s}L_{\rm m}\label{eqRm}\\
{\rm Pm} = 2\ 10^{17} \displaystyle\frac{T_{e,\rm eV}^{3/2}T_{i,\rm eV}^{5/2}}{n_{\rm m^{-3}}\sqrt{M_{\rm amu}}}\label{eqPm}
\end{eqnarray} 
$U$ being the typical plasma flow velocity (expressed in km/s), $L$ being the characteristic size (expressed in m) and $M_{\rm amu}$ the ion mass (expressed in atomic mass unit).
Given the parameters in table~\ref{tab:plasmaparam}, the maximum expected magnetic Reynolds number is of order 10, while Pm can be varied from $10^{-5}$ to 1. These parameters will thus allow to study regimes for which magnetic induction due to the velocity field is weakly non-linear. The wide range of accessible Pm values gives a unique opportunity to bridge regimes obtained experimentally in liquid metals (for $\mbox{Pm}\sim 10^{-6}$) to regimes currently reached in numerical simulations (in the range $10^{-2}$ to 1).

Note that corrections due to magnetization occur when $\omega_{ci}\tau_{ii}\neq 0$~\citep{Braginskii1965}, where $\tau_{ii}$ is the ion-ion collision time (for a singly charged ion $\tau_{ii, \rm s} = 3\ 10^{13}\sqrt{\mu/2}T_{i, \rm eV}^{3/2}/\Lambda\ n_{\rm m^{-3}}$,  $\Lambda$ being the Coulomb logarithm). For the typical parameters presented in table~\ref{tab:plasmaparam}, $\omega_{ci}\tau_{ii}$ will be in the range 0.1 (for the highest densities, lowest magnetic fields) to 40 (for the lowest densities, highest magnetic fields) and corrections have to be taken into account; this is one of the addressed questions detailed below.

Equations~(\ref{eqRm}) and~(\ref{eqPm}) show that, for given electron and ion temperatures, Rm and Pm are controlled by the flow intensity and the plasma density respectively, which are independently controlled in our device. One should however note that the plasma velocity may be bounded by a maximum value close to the Critical Ionization Velocity, as observed in~\cite{Collins2014} for similar plasma parameters --- in this configuration the plasma density is not only controlled by the radio-frequency power but also depends on the cathode's current.

\begin{table}
\begin{center}
\def~{\hphantom{0}}
  \begin{tabular}{lllll}
      Parameter  & Symbol & Current device  &  Final device  & Units\\[3pt]
       Plasma density& $n$   & $10^{16}$ - $5.10^{18}$ & $10^{16}$ - $5.10^{19}$ & m$^{-3}$\\ 
       Electron Temperature & $T_e$  & 2 - 6 & 2-10  & eV\\
       Ion Temperature  & $T_i$ &0.1 - 1 & 0.1 - 1 & eV\\
       Ionization fraction & $f_\%$   & 0.1 - 30 & 0.1 - 90 & \%\\
       Magnetic field & $B_0$ & 5 - 200 & 5 - 200 & mT\\
       Flow velocity &	$U$	& 0 - 3  & 0 - 10 & km.s$^{-1}$\\
  \end{tabular}
  \caption{Typical plasma parameters in the current stage experimental device and the final projected setup}
  \label{tab:plasmaparam}
  \end{center}
\end{table}


\subsection{Plasma diagnostics}
Various plasma diagnostics tools are being developed to probe the plasma parameters. The first set of diagnostics deals with intrusive probes, the relatively low plasma densities and temperatures allowing for long-time operation. A swept radio-frequency compensated Langmuir probe~\citep{Cantin} gives access to plasma density $n$, electron temperature $T_e$, plasma potential $V_p$ and floating potential $V_f$ using standard analysis techniques~\citep{Hershkowitz}. Emissive probes are also used to probe plasma potential~\citep{Sheehan}. Plasma flows are measured using two-face Mach probes biased negatively (at least $5T_e$ below plasma potential) to collect ion saturation current (see figure~\ref{figure:Expschematic}). Mach probe data are analysed using standard techniques~\citep{Chung}.
A miniature two-face retarding field energy analyser (RFEA) is currently under development and will allow the determination of ion flux and energy distribution functions in flowing plasmas. The analyser combines electrostatic grids and magnetic barriers for electron suppression. In classical RFEAs with 4 grids, the effective transparency is a function of the discriminator grid bias due to lens effects within the grids and the measured energy distribution is broadened by instrumental effects. With the new design, only one electrostatic grid is needed within the analyser and the transparency for ions is constant over the measured ion energy range and also allows to probe the global spatio-temporal dynamics of plasma density fluctuations~\citep{Rafalskyi}.

Magnetic fields will be measured using arrays of Hall probes and high frequency magnetic fluctuations will be measured with Rogowski probes in probe shafts. All these electrical and magnetic diagnostics probes are mounted on motorised linear translation stages and give access to spatially resolved profiles of the plasma parameters.
 
Optical diagnostics are also being implemented in the device. Optical Emission Spectroscopy uses plasma light emission to determine the details of the nature of the ionized and excited species, and may be used to measure the electron temperature and density. A Ocean Optics USB2000+ spectrometer gives access to the emitted spectral lines and will be used to independently measure the electron temperature. Preliminary results related to the dynamics of low-frequency waves at the edge of the plasma column have been obtained with a high-speed camera (Photron SA-5). High-speed imaging (around $10^5$ images/s at a resolution 200x200 px) allows to probe the global spatio-temporal dynamics of plasma density fluctuations~\citep{Oldenburger}.
Ion velocimetry from Laser Induced Fluorescence of Argon ions (ArII) lines will shortly be installed to probe the details of the plasma flows with non-intrusive diagnostics~\citep{Demtroder, Bieber}.


\subsection{Numerical simulations}
In parallel to the experimental development, numerical simulation have been carried out in order to guide our understanding of the flow dynamics and of the flow drive configuration. In the following, we present results obtained in the "one-cell" configuration with two different flow drive geometries. We consider the dynamics of an incompressible magnetofluid within the MHD description. Effects of compressibility, inhomogeneous density, temperature and charge distributions are not taken into account at present stage, but it is shown that already this level of complexity reveals a wealth of interesting phenomena. In this study we will in particular focus on plasma flow driving schemes and how the induced plasma velocity depends on the different physical parameters. We consider a uniform plasma in a cylindrical vessel of radius $R$ and length $L$, as a simplified model of the plasma column observed in the experimental device. The imposed magnetic configuration corresponds to a uniform, stationary axial magnetic field on which we superimpose a radial current between two electrodes (Figure \ref{Fig:forcing}), mimicking the one-cell experimental configuration. We have numerically tested two schemes for driving flows from an azimuthal $\bm J \times \bm B$ torque. The first one consists of two point-electrodes, generating a linear current. The second one consists of two concentric circular electrodes in a plane perpendicular to the axis, generating a radial current in the annular space between the 
electrodes. This force is applied to a small, linearly extended domain in the case of the two point-electrodes. In the case of the circular electrodes the torque acts over the annular domain between the two circles.

The equations we consider, in conveniently normalized Alfv\'enic units are,
\begin{eqnarray}
 \frac{\partial \bm u}{\partial t}+\bm u\cdot \nabla \bm u=-\nabla p+Pm S^{-1} \Delta \bm u+\bm J\times\bm B,\label{eq:u}\\
 \frac{\partial \bm B}{\partial t}=\nabla \times (\bm u \times \bm B)+S^{-1} \Delta \bm B\label{eq:B},\\
\nabla \cdot \bm u=0~~~~\nabla \cdot \bm B=0.
\end{eqnarray}
In these equations $\bm u$ is the velocity, $\bm B$ the magnetic field,  $p$ the pressure normalized by the density, and $\bm J =\nabla\times \bm B$ the current density. Once the geometry of the domain and of the forcing is fixed, the dynamics of the system of equations we consider is fully determined by three parameters:  the Lundquist number $S=B_0 L/\eta_B$, the magnetic Prandtl number $Pm=\eta_u/(\rho\eta_B)$ and the forcing strength $J_0 L/B_0$, with  $B_0$ the axially imposed magnetic field. The Lundquist number is defined similarly to the magnetic Reynolds number, with the Alfven speed chosen as the characteristic velocity (Rm being then an output of the simulation with the real obtained flow velocity). The reference length is the cylinder length $L$ and $J_0$ is the current density flowing between the two electrodes or the radial current density at the inner circular diode. In order to compare the two forcing schemes, we have fixed the same total amount of current flowing between the electrodes.

 Equations (\ref{eq:u}) and (\ref{eq:B}) are discretized with a three-dimensional Fourier pseudo-spectral method on a Cartesian grid. To impose the boundary conditions we use the volume-penalization technique, a method of the immersed boundary type. The method, previously succesfully used to investigate the influence of geometry on toroidally shaped plasma \citep{Morales2012} is presented in detail for three-dimensional viscoresistive MHD equations in \cite{Morales2014}. We shall only briefly outline its main features. 

The numerical domain is a rectangular box of size $4/5\pi R\times4/5\pi R\times 4\pi R$. Within this domain, we define a cylindrical sub-domain of length $L = 10 R$ and radius $R$. We consider the plasma to be confined within this closed cylinder while the remainder of the rectangular box  represents solid walls. The forcing is placed at a distance $1.85 R$  away from the upper wall. The linear current forcing is obtained by imposing a current to flow between electrodes at $r=0$ and $r=0.7 R$. The circular electrodes are placed at $r=0.1R$ and $r=R$.

The MHD equations are solved in the entire numerical domain, but the velocity tends rapidly to zero outside the cylindrical vessel 
where the volume penalization term becomes active. The magnetic field is not directly influenced by the plasma-wall interface and 
satisfies the periodic boundary conditions on the boundary of the numerical domain. In practice the impact of this simplification is reasonably small since the induced magnetic field fluctuations are small outside the cylindrical vessel where the velocity is practically zero. The simulations are carried out in a cubic domain with $128^3$ grid points for all simulations presented in this  paper. Time-advancement is performed by a third-order Adams-Bashforth scheme, from zero velocity initial conditions. 


\section{A facility to study plasma flows and magnetic induction}\label{sec:3}


\subsection{Magnetic induction processes at variable Pm}
The von-K\'arm\'an plasma experiment has been designed as a versatile device for driving several types of axisymmetric flows in a cylinder, with a maximum magnetic Reynolds number of order 10. These flows are thus ideally suited for the study of magnetic induction in the presence of an externally applied magnetic field. 
As an example, a $s_2t_2$ flow displaying a strong azimuthal shear layer, efficiently induces an azimuthal magnetic component from an axial magnetic field. The von-K\'arm\'an plasma experiment will allow to study these induction processes both as a function of Rm and Pm, extending the  results obtained at low Pm values in liquid metals~\citep{VerhilleSSR2010}. The independent control of Rm and Pm will allow to investigate the influence of Pm (modifying the plasma density $n$) --- at constant Rm (keeping the driving torque constant) --- on the dynamics of basic induction processes. The versatility of the von-K\'arm\'an plasma experiment will allow to probe several magnetic induction mechanisms in the presence of time-averaged plasma flows with one or two poloidal and/or poloidal cells. Departures from the simple MHD framework are expected in this device, and their influence on the magnetic induction processes will indeed be addressed.


\subsection{Plasma fluctuations in the presence of large-scale driven flows}
Driving intense flows with typical Rm values of order 10 in a range of Re from 10 to $10^5$ will give the opportunity to probe the joint dynamics of velocity and magnetic field fluctuations. Two distinct issues will be addressed:
\begin{itemize}
\item Spectral analysis and kinematic-magnetic cross-correlations. Higher orders than two-point statistics (magnetic and velocity increments) will be investigated to address scale dependent statistical behaviours as classically observed in neutral turbulence (e.g. intermittency phenomenon). Intrinsic values of the dimensionless parameters will also be measured. Note that Re and Tm are not uniquely determined, as there are multiple choices for the typical scales. It is common, in neutral turbulence studies, to define an objective Reynolds number Re$^i$, based on intrinsic characteristic scales of the turbulence. This is generally done by considering, for the typical velocity, the root mean square $U_{\rm rms}$ of the turbulent fluctuations, and, for the length scale, either the injection energy integral scale $L_i$ (correlation length of the turbulent velocity field), or the Taylor microscale $\lambda_U$ (ratio of the velocity fluctuations to the fluctuations of the velocity gradients). These scales are not anymore input parameters of the problem, but output, as they can only be determined from precise measurements of the turbulent fields. We propose here to follow the same guideline as in neutral turbulence in order to determine intrinsic Re$^i$ and Rm$^i$~\citep{Weygand}. From these intrinsic Reynolds numbers, we get a direct measure of the extent of the inertial range of the kinematic and magnetic turbulent cascades. For instance, an interesting point will be to consider their ratio as an effective magnetic Prandtl number of the flow, Pm$^i$, and to investigate how it evolves with Pm determined from transport coefficients. Pm and Pm$^i$ are expected to be simply related only if the kinematic and magnetic scales in the definition of the corresponding Reynolds numbers are themselves identical or simply related.
\item Influence of intense rotation on low-frequency fluctuations dynamics. The large plasma pressure and plasma potential gradients at the edge of the plasma column perpendicular to the axial magnetic field are sources of free energy to excite low frequency instabilities (drift waves or centrifugal instability). The existence of azimuthal shear may also be Kelvin Helmholtz unstable. The current version of our experimental setup (in the one-cell configuration) will allow to explore the influence of intense global rotation on the growth and dynamics of these low-frequency modes. In particular, the role of large-scale controlled flows on effective small-scale cross-field transport can be addressed. Fast imaging camera and probe-arrays will be used to probe the spatio-temporal dynamics of plasma fluctuations~\citep{Oldenburger}.
\end{itemize}


\subsection{Flow drive mechanisms and momentum transport in partially ionised weakly magnetised plasmas}
The typical dimensionless parameters given in the previous section were computed for an unmagnetized plasma (or, equivalently, in the parallel direction). However, we showed that corrections due to magnetization have to be taken into account. Measurement of the spatial evolution of the plasma flow (in the radial and axial directions) will allow to determine the components of the viscosity tensor, and relate their scalings with the evolution of plasma parameters (since the plasma viscosity is expected to vary with the plasma density $n$ and electron temperature $T_e$) and the amplitude of the magnetic field. In partially ionised plasmas, the dynamics of the charged particles may be dominated by collisions with neutrals~\citep{Lieberman}; in conditions similar to our setup, it has already been reported that viscous transport may strongly be affected by neutral drag (typically from charge exchange collisions of ions with neutrals)~\citep{Collins2012,Collins2014}.
The detailed analysis of the evolution of transport coefficients as a function of the magnetisation parameter $\omega_{ci}\tau_{ii}$ is of key importance in the modelling of dynamo saturation mechanisms. The dynamo instability stems from an unmagnetised medium in the presence of intense motions (for which the transport coefficient are scalar quantities). Saturation is obtained when the magnetic field has grown to values high enough for the Lorentz force to modify the flow (typically to equipartition). In this saturated regime, the dynamo magnetic field introduces an anisotropy, and the transport coefficients are tensor quantities.

Our experimental setup will also provide a benchmark for plasma acceleration mechanisms. This benchmark will be instrumental in the development of future plasma dynamo experiments (or other related dedicated plasma experiment for studying processes in flowing plasmas) or applications in the space propulsion context.


\section{Preliminary results on plasma flow drive}\label{sec:4}


\subsection{Experimental results}

\begin{figure}
\centering\includegraphics[width = \linewidth,angle=0]{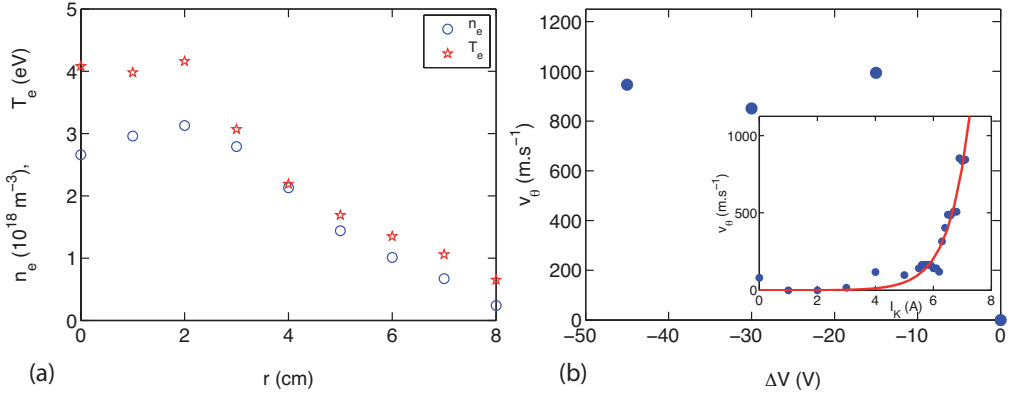}
\caption{(a) Radial profiles of plasma density ($\circ$) and electron temperature ($\star$). (b) Evolution of orthoradial velocity as a function of cathode-anode voltage (for a 3 A emited current, $I_K = 7$~A). Inset: Evolution of orthoradial velocity as a function of heating current $I_K$. 750 W r.f. power and 30 mT applied magnetic field, 0.3 Pa in Argon.}
\label{Fig:expresults}
\end{figure}

Results presented in this subsection have been obtained with the one-cell configuration experimental setup as described in figure~\ref{figure:Expschematic}. 
In this configuration, a 7 cm diameter plasma of density of peak density  $3.10^{18}$~m$^{-3}$ and electron temperature 4 eV is produced. The radial profiles of plasma density and electron temperatures as measured from the swept Langmuir probe are displayed in figure~\ref{Fig:expresults}(a), showing the extent of the high density plasma column up to $r\sim3.5$ cm.  The flow driving method described above has been investigating in regimes where the plasma is stable to low frequency fluctuations (drift wave, centrifugal instability of KH modes) and no fluctuations of the plasma parameters have been measured.


The emissive cathode made from 0.25 mm in diameter Thoriated Tungsten filament has been inserted at radial location $r\sim 5$ cm (i.e. slightly outside the high density plasma column). The cathode is heated by circulation of a DC current of intensity $I_K$. It has been checked that the emitted current increases exponentially with the heating current $I_K$ (accordingly to the Richardson law of thermo-electronic emission) above a threshold, and when the cathode is sufficiently negatively biased. The emissive cathode is negatively biased at a potential $\Delta V$ relative to a Tungsten anode located at $r=0$ and at the same axial location. The cathode/anode assembly is left floating so that the net current of the assembly is null. The flow velocity has been measured by the two-face Mach probe. The ratio $R_M$ of the ion saturation currents from the two faces is related to the plasma velocity as $v = C_s \ln{R_M}/1.34$~\citep{Chung}, where $C_s$ is the Bohm velocity (or sound speed). 
Figure~\ref{Fig:expresults}(b) shows the evolution of the ion flow velocity  as a function of $\Delta V$ or an emitted current of amplitude 3 A. This graph shows that, for a given filament temperature of the cathode, there is no flow when the cathode potential is identical to the anode potential. A flow reaching 900 m/s is observed as soon as $\Delta V<-15V$, i.e. when the cathode is biased sufficient negatively to emit electrons. The observed flow velocity is in the direction of the $\bm J \times \bm B$ torque. The inset of figure~\ref{Fig:expresults}(b) shows the evolution of the velocity when $I_K$ (the current heating the filament) is varied, keeping $\Delta V = -15V$. From this evolution, it is clear that the measured velocity decreases as $I_K$ decreases. The red curve shown in the graph being proportional to the exponential fit of the emitted current. This shows that the measured plasma velocity is proportional to the emitted current and controlled by the amplitude of the $\bm J \times \bm B$ torque

A strong limitation of the present setup lies in the high sputtering rate of the tungsten filament, which leads to metallic coating on Mach probes and limited operation time. The use of the LaB$_6$ cathode, just assembled at the time of writing and shown in figure~\ref{figure:Expschematic}, should allow for longer time operation and measurement of spatial profiles of velocity.

\subsection{Numerical results}

\begin{figure}
\includegraphics[width=0.5\linewidth,angle=0]{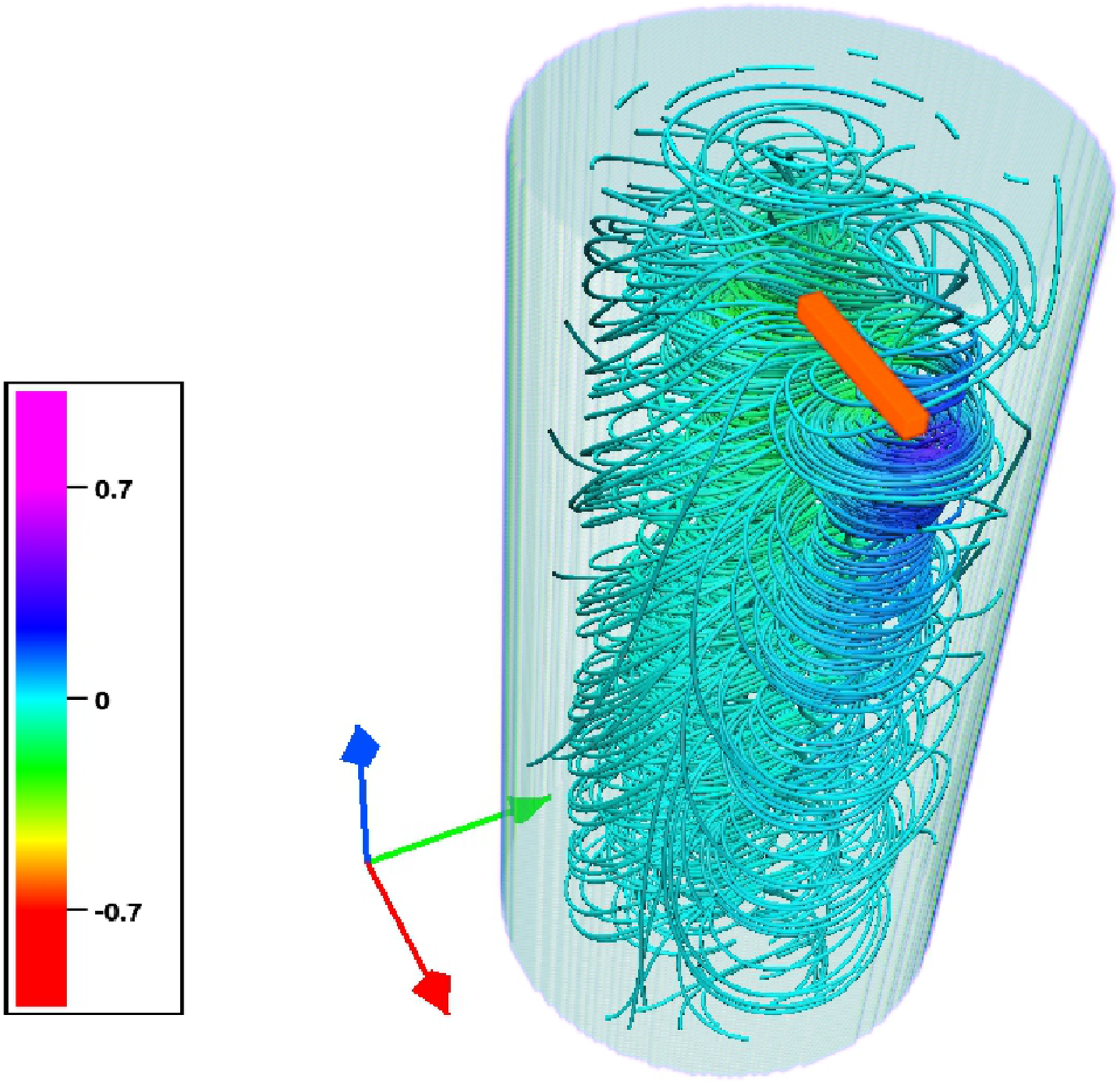}~
\includegraphics[width=0.47\linewidth,angle=0]{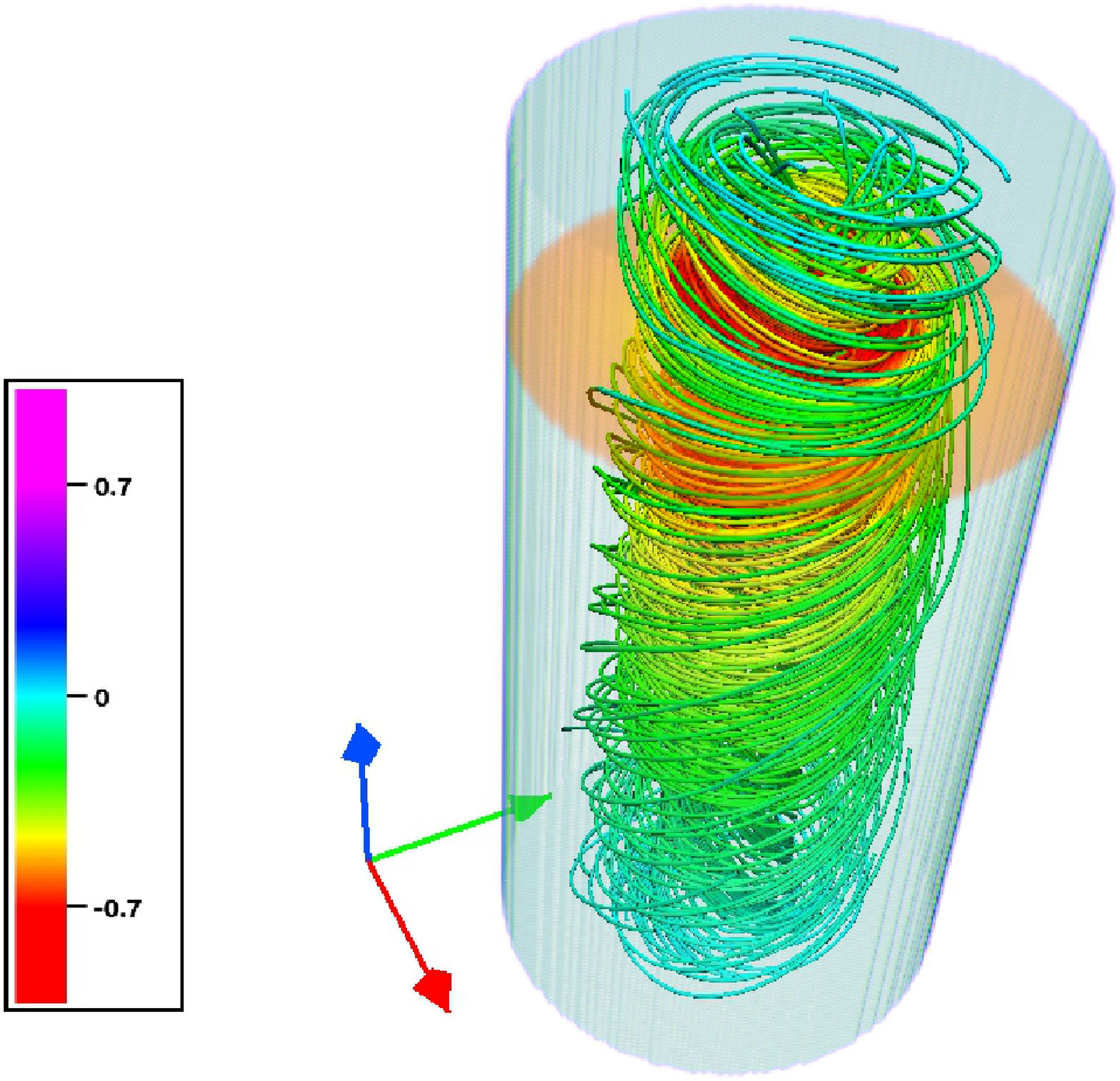}
\caption{Numerical simulation of two flow drive configurations. The first one (left) consists of two point-electrodes, generating a linear current. The second forcing (right) consists of two concentric circular electrodes, generating a radial current in the annular space between the electrodes. Displayed are streamlines, coloured with the value of the azimuthal velocity (in Alfvenic units).}
\label{Fig:forcing}
\end{figure}

 All results we present correspond to the steady state to which the system was observed to evolve, after the axially propagating Alfv\'en waves generated by the impulsively started forcing current had died away. We have fixed the forcing strength and varied the Lundquist and Prandtl numbers in the ranges $1\leq S \leq 1000$, $0.001\leq Pm \leq 1000$.

\begin{figure}
\includegraphics[width=0.5\linewidth,angle=0]{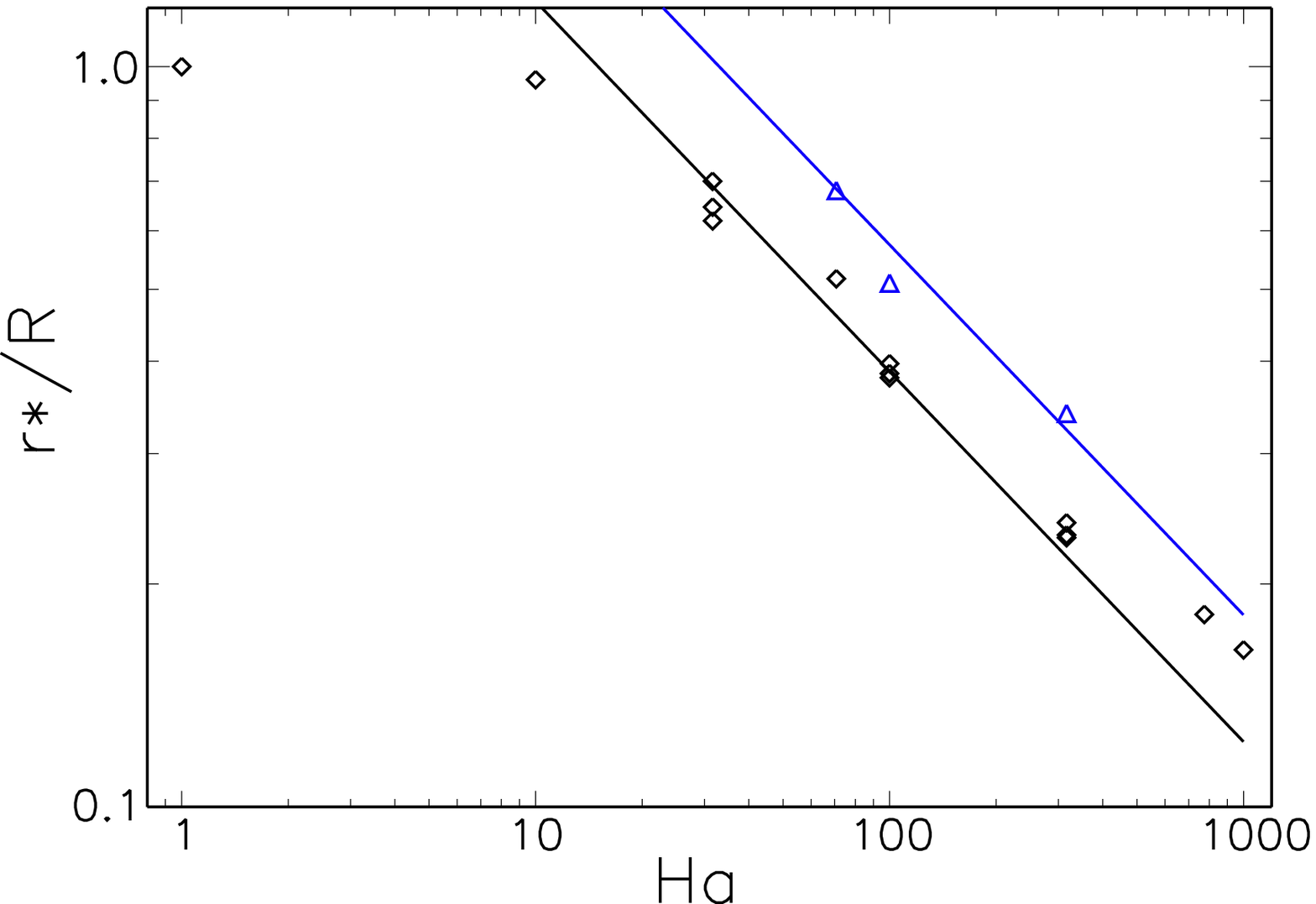}~
\includegraphics[width=0.5\linewidth,angle=0]{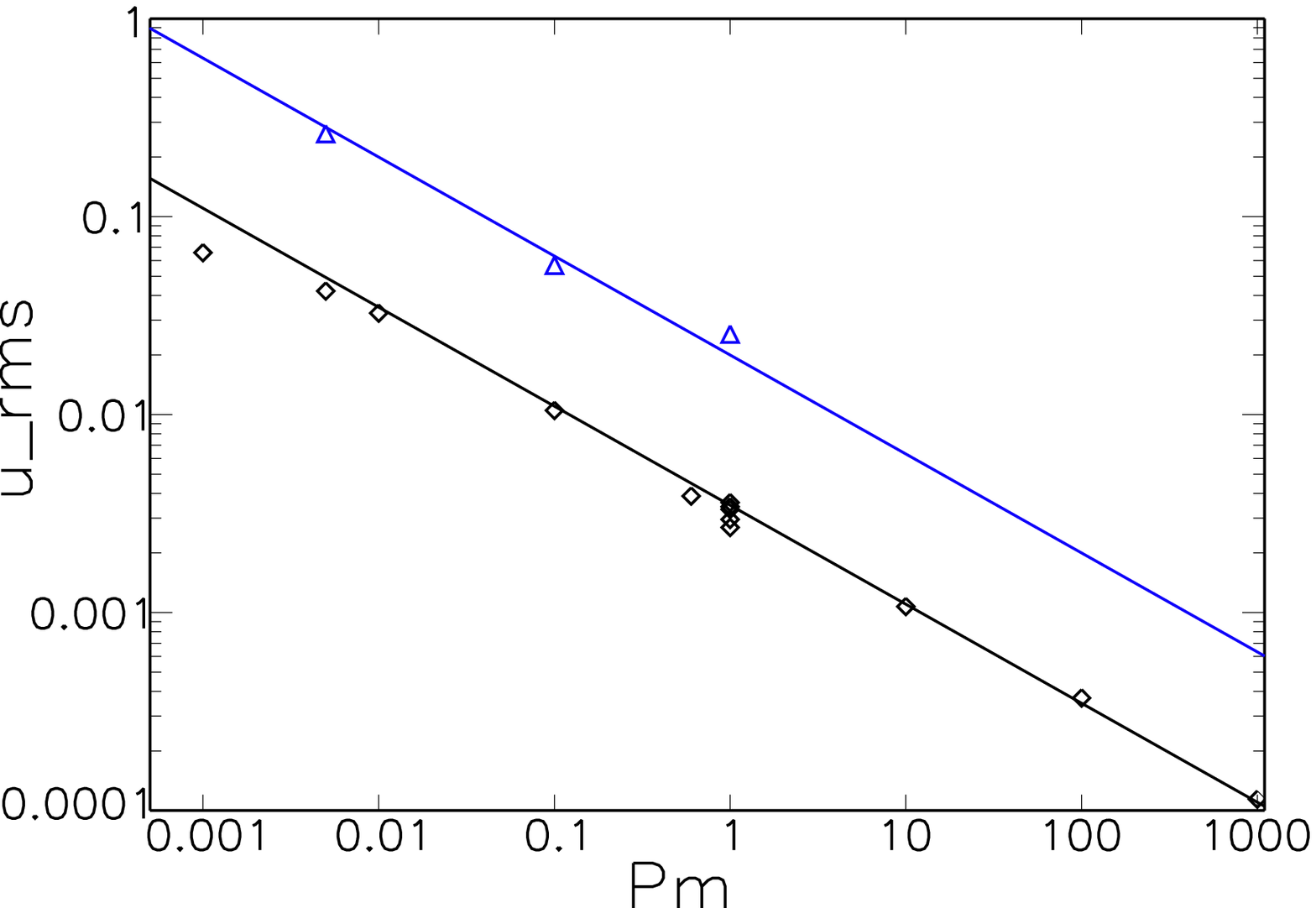}
\caption{Left: radius of the vortical structure induced by the two forcing protocols as a function of $Ha$. Right: RMS value of the velocity in the plasma (in Alfv\'en units) as a function of $Pm$. Black lozenges corresponding to the linear forcing (see figure~\ref{Fig:forcing}(left)) and blue triangles corresponding to the annular forcing (see figure~\ref{Fig:forcing}(right))}
\label{Fig:radius}
\end{figure}

We observe that, depending on the value of $S$ and $Pm$, a part of the plasma is set into circular movement. However, depending on the value of the different parameters, the radius of the azimuthally moving plasma column can vary importantly. In the limit of vanishingly small magnetic Prandtl number, {\it i.e.} in the hydrodynamic limit, this column radius is limited by the size of the cylindrical domain.  We have quantified the dependence of the radius of the vortical structure by determining its value as 
\begin{equation}
 r^*=\frac{\int_0^R  \int_0^{2\pi} u_\theta(r)r^2 d\theta dr}{\int_0^R \int_0^{2\pi} u_\theta(r)rd\theta dr},
\end{equation}
where we have evaluated $u_\theta(r)$ in the plane of the forcing. Figure \ref{Fig:radius} (left) shows the results for the radius of the vortical structure induced by the two forcing protocols (see figure~\ref{Fig:forcing}) as a function of the Hartmann number, 
\begin{equation}
 Ha=S~Pm^{-1/2}.
\end{equation}
It seems that it is the Hartmann number which determines the size of the vortical structure, roughly independent of the value of the magnetic Prandtl number with a dependence proportional to $Ha^{-1/2}$. At low values of $Ha$, the size of the vortex is limited by the size of the domain. We also see from this figure that the two different forcing methods behave qualitatively the same, the circular electrodes generating a larger vortical structure for the same parameters. 

In order to quantify how efficiently the forcing can set the plasma into movement we also measure the root-mean-square value of the total velocity, averaged over the fluid domain. The results are reported in Figure \ref{Fig:radius} (right). There is clearly a decreasing trend of the plasma velocity as a function of the Prandtl number and all results approximately collapse on one line with slope $Pm^{-1/2}$. In particular, it is observed that for the same current and the same values of the dimensionless numbers, the circular forcing induces an RMS value almost one order of magnitude larger than the linear forcing. Part of this is created by the radial extension of the circular forcing which is larger than that of the linear forcing. The details of the above scalings will be reported in a future investigation.

\section{Conclusions}

We have introduced a novel versatile experimental device designed to explore the dynamics of basic magnetic induction processes and the dynamics of flows driven in weakly magnetized plasmas. The expected final experiment will drive a von-K\'arm\'an like flow and will be based on two radio-frequency plasma sources at each end of a 1 m long linear device create a high-density plasma column ($10^{16}$--$10^{19}$  particles.m$^{-3}$). Flows are driven through $\bm J \times \bm B$ azimuthal torques created from independently controlled emissive cathodes. The device has been designed such that magnetic induction processes and turbulent plasma dynamics can be studied in a variety of time-averaged axisymmetric flows in a cylinder. In the first stage of the project, flow drive mechanisms have been tested in one half of the expected device, where only one plasma source and one cathode have been implemented. Preliminary experimental results show that a rotating motion of up to nearly 1 km/s is controlled by the $\bm J \times \bm B$ azimuthal torque.These experimental investigations are supported by MHD simulations implementing volume-penalization, which were carried out in order to design the most efficient flow-driving schemes and understand the flow dynamics.
This experimental device will allow to address the influence of the magnetic Reynolds number Rm and magnetic Prandtl number Pm on the joint velocity field - magnetic field dynamics. It has been designed to cover a broad range of Pm values, from very low values (similarly to liquid metals) to values of order 1. In particular, the following issues will be detailed:
\begin{itemize}
\item Influence of Pm on magnetic induction. In particular, we will investigate the influence of the scale separation of the dissipative scales for the velocity and the magnetic fields on magnetic induction. Departure from the MHD framework will also be investigated.
\item Plasma fluctuations in the presence of large scale driven flows. Intrinsic values of the dimensionless parameters and their evolution with the plasma parameters, and their comparison with theoretical computation from transport coefficients will be analysed. The influence of intense rotation on low-frequency instabilities and cross-field transport will also be investigated
\item Flow drive mechanisms. A preliminary task of the project will be to efficiently drive flow and characterise how the plasma responds to flow forcing. In particular, transport coefficients (and possible corrections due to magnetisation) will be experimentally determined as a function of the plasma parameters.
\end{itemize}
\vspace{1 cm}

\textbf{Acknowledgements}
This work has been supported by French National Research Agency under contract ANR-13-JS04-0003-01, and by the LABEX iMUST (ANR-10-LABX-0064) of Universit\'e de Lyon, within the program "Investissements d'Avenir" (ANR-11-IDEX-0007) operated by the French National Research Agency. CPU time for the numerical simulations was provided by IDRIS under project number 22206. N. Plihon thanks Laboratoire de Physique for supporting the initial development of the project, F. Debray at LNCMI for providing the Bitter coils, J. Guillon for numerous technical advices and P. Chabert at LPP for fruitful discussions and providing the r.f. generator. Fruitful discussions with the group of Pr Forest at UW Madison are acknowledged.

\bibliographystyle{jpp}

\end{document}